\documentclass[reprint, 
superscriptaddress,
amsmath,amssymb,
aps,
prfluids,
floatfix]{revtex4-2}

\usepackage{graphicx}
\usepackage[T1]{fontenc}
\usepackage{soul}
\usepackage{newtxtext}
\usepackage{newtxmath}
\usepackage[
    colorlinks,
    citecolor=blue,
    urlcolor=blue,
    linkcolor=blue
    ]{hyperref}
\usepackage{orcidlink}

\newcommand{\reynolds}{\textrm{Re}}
\newcommand{\froude}{\textrm{Fr}}

\begin{document}
\title{Preventing sinking of a disk by leveraging the boundary jump phenomenon}

\author{Jan Turczynowicz\orcidlink{0009-0002-8745-2298}}
\affiliation{Fenix Science Club, {Aleja Stanów Zjednoczonych 24, 03-964 Warsaw}}
\affiliation{Institute of Theoretical Physics, Faculty of Physics, 
University of Warsaw, Pasteura 5, 02-093 Warsaw, Poland}
\author{Radost Waszkiewicz\orcidlink{0000-0002-0376-1708}}
\affiliation{Institute of Physics, Polish Academy of Sciences, Aleja Lotników 32/46, PL-02668 Warsaw, Poland}
\affiliation{Fenix Science Club, {Aleja Stanów Zjednoczonych 24, 03-964 Warsaw}}
\author{Łukasz Gładczuk\orcidlink{0000-0001-7252-7619}}
\email{lukasz@fenix.club}
\affiliation{Fenix Science Club, {Aleja Stanów Zjednoczonych 24, 03-964 Warsaw}}

\date{\today}

\begin{abstract}
Although it is commonly expected that a metal disk placed on the surface of water will sink, our investigation has revealed a surprising phenomenon: a vertical jet directed onto the disk from above can allow it to remain afloat. 
This result defies intuition, as one would assume that the force of the jet's impact would cause the disk to sink.
We have discovered that this phenomenon occurs as a result of water displacement from the top of the disk caused by the impacting jet, operating through a mechanism similar to the hydraulic jump. 
This displacement increases the effective immersed volume, resulting in an increased buoyant force that balances gravity.
In contrast to the classical case, here the jump radius is fixed by the geometric parameters of a disk, a phenomenon we refer to as the \emph{boundary jump}.
To further explore this effect, we have presented a theoretical model based on scaling laws, which provides the conditions required for the disk to float. 
The prefactor was determined through an independent experiment.
Finally, we conducted experiments on the disk's floating and sinking, which showed a good match with the proposed theory.
\end{abstract}
                              
\maketitle

\section{Introduction}

\begin{figure}
\begin{center}
\includegraphics[width=1\linewidth]{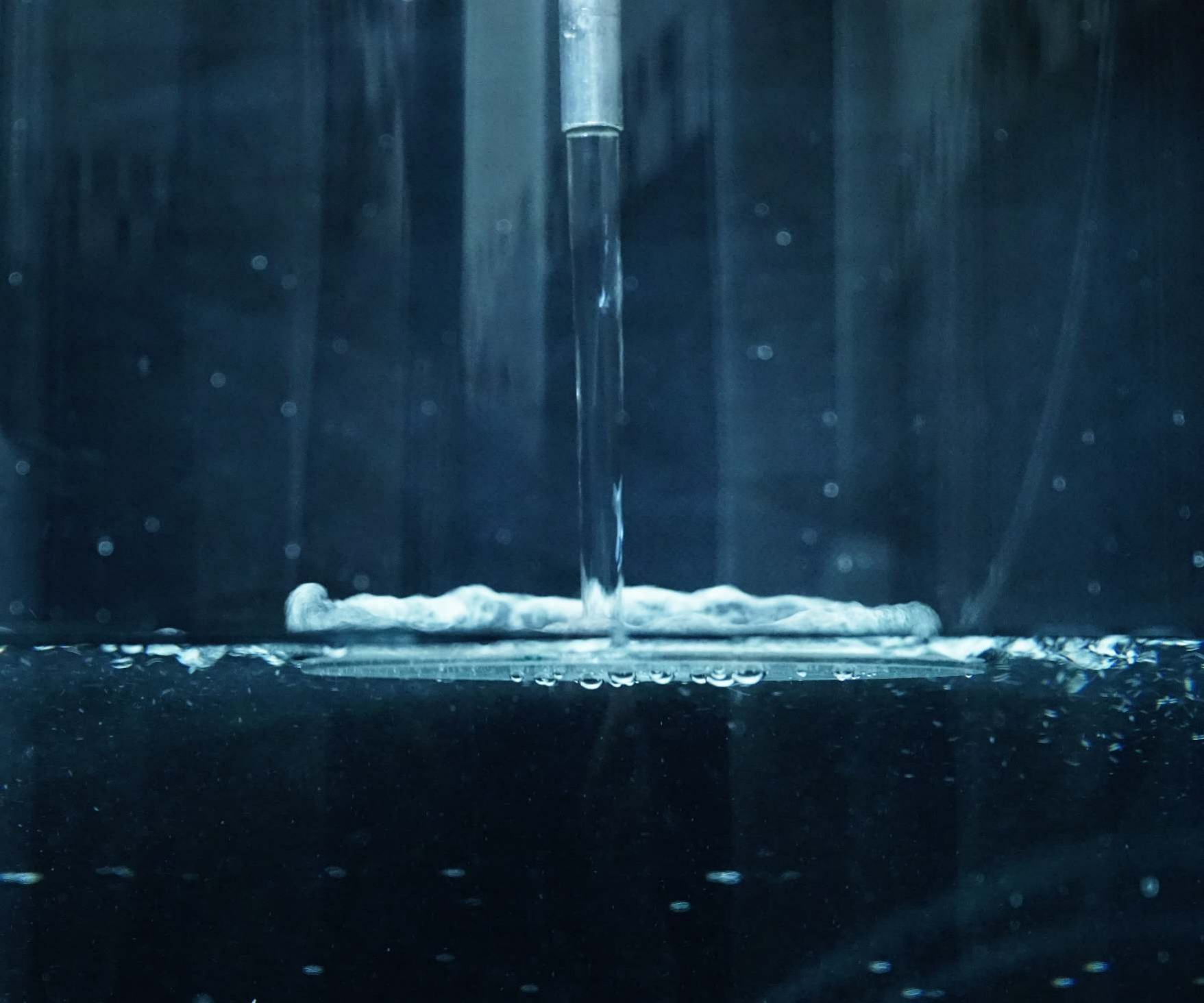}
\end{center}
\caption{A side view of a metal disk floating just below the surface of the water, with a boundary jump occurring at the edge of the disk. The photograph was taken in a transparent cylindrical vessel for demonstration purposes only. Due to a lensing effect caused by the vessel, the disk appears larger than the surface disturbance. A video capturing this phenomenon can be viewed at the following link: \url{https://youtu.be/as0wRQj1Zws}.
}\label{fig:parametric_jumps_x}
\end{figure}

Placing a thin metal disk on the surface of water and directing a vertical water jet onto the disk reveals an unusual phenomenon. Despite the disk being denser than water, in certain cases, the weight of the water displaced from the disk’s surface is sufficient for the buoyancy force to balance both the impact force of the jet and the weight of the disk. This balance allows the disk to remain afloat as shown in Figure~\ref{fig:parametric_jumps_x}.

To understand the conditions necessary for this phenomenon to occur, it is essential to study the water flow dynamics on the upper surface of the disk. Due to the cylindrical symmetry of the flow and the very thin layer of fast-moving liquid, this process is reminiscent of the classical hydraulic jump problem. The hydraulic jump has been a subject of research since at least the works of \citet{Rayleigh_1914}. Today, despite different variants of the flow geometry being studied \citep{Ivanova_2019,Teymourtash_2015,Wang_2019,Wang_2018}, details of the phenomenon remain a subject of interest \citep{Baayoun_2022,Duchesne_2022}.

Qualitatively, when a jet strikes the plate, a region of higher pressure necessarily forms at the center of the disk. In the region very close to the disk's center, where the boundary layer flow is not yet developed \citep{Baayoun_2022}, the flow pattern resembles that of an inviscid flow, where high pressure at the center accelerates the liquid outward. As the fluid moves away from the center, a combination of outward acceleration and increased circumference leads to a rapid decrease in film thickness. When the boundary layer reaches the surface, viscosity dominates; however, due to the increasing circumference, the film thickness remains small.

At some distance $R_{\mathrm{J}}$ from the center, the thin layer becomes unstable, leading to an abrupt transition to a subcritical, thicker layer. Many studies \citep{Watson_1964, Bohr_1993, Bohr_2021, Duchesne_2019} have investigated this phenomenon and analyzed the value of $R_{\mathrm{J}}$. Generally, $R_\mathrm{J}$ increases with jet flow $Q$, but for the finite disk of radius $R$, this relationship only holds when $R_{\mathrm{J}} < R$. When $Q$ reaches values that would cause the radius of the jump to exceed the disk’s radius, the jump position becomes fixed at $R$.

\begin{figure*}
\begin{center}
\includegraphics[width=1\linewidth]{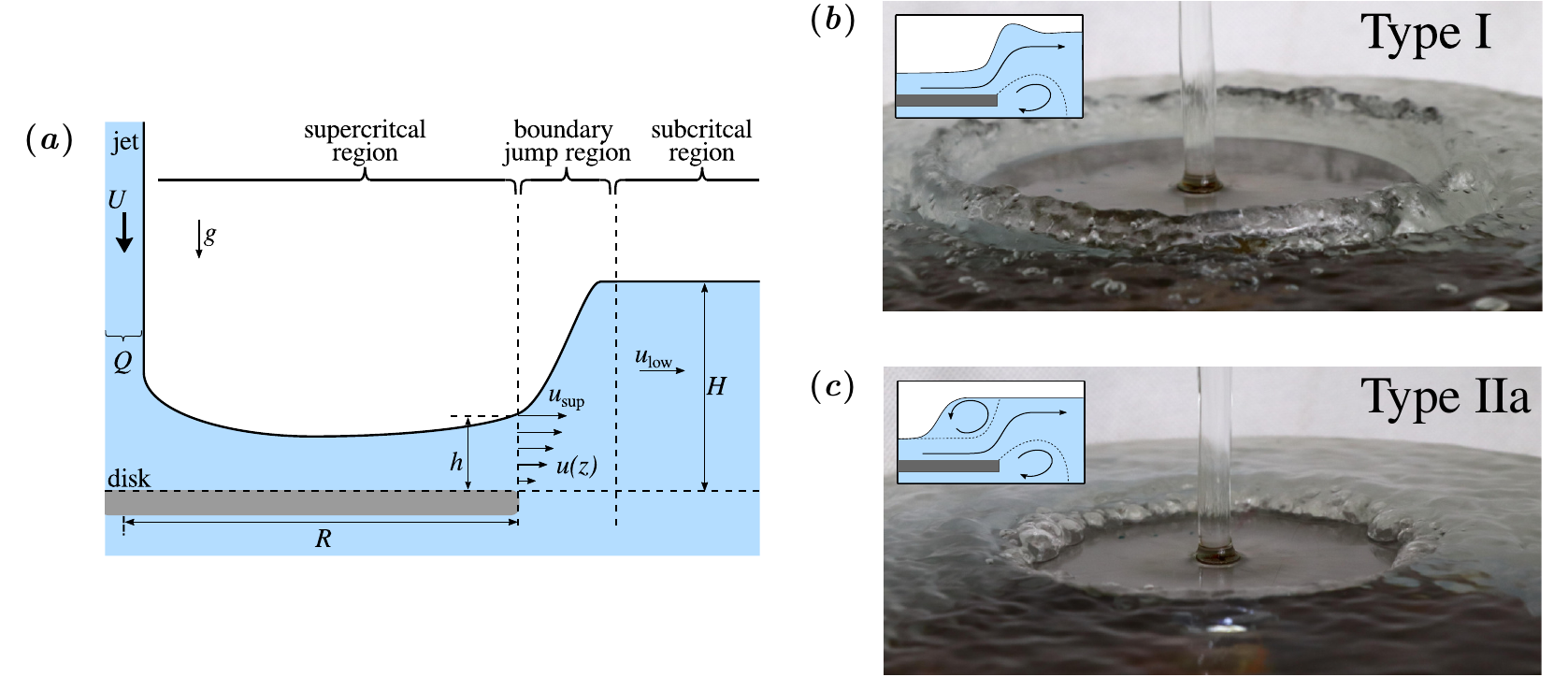}
\end{center}
\caption{
Sketch (a) shows the flow pattern on the top of the disk. The jet impacts the disk, creating a supercritical layer. Near the disk's edge, this layer has a thickness $h$ and a surface velocity $u_{\textrm{sup}}$. At the edge the boundary jump occurs, with a rapid increase in the liquid layer thickness to $H$ and a reduction in liquid velocity to $u_{\textrm{low}}$. Depending on the value of $H$ two types of jumps are observed. At $H= 2.5\,\mathrm{mm}$ a type I jump (b) occurs -- where the entire disk is visible with a characteristic high wave formed close to its edge. For a larger value of $H =5.5\,\mathrm{mm}$ a type IIa jump (c) is observed -- the edge of the disk is covered, and water close to the jump edge is moving \emph{towards} the centre of the disk. The experiments shown in photos (b) and (c) used disks with $R = 6\,\mathrm{cm}$ and jet with $Q = 104\,\mathrm{ml/s}$ and $U = 1.7 \,\mathrm{m/s}$.
}\label{fig:parametric_jumps}
\end{figure*}

When the moving liquid encounters the stationary liquid outside the disk, its movement is abruptly halted, causing the fluid level to rise. Experimental investigations revealed that, as the disk’s submersion depth increases, a transition occurs between two flow patterns (Figure~\ref{fig:parametric_jumps}b-c). We refer to these regimes as type I and type IIa jumps, following the nomenclature of \citet{Ellegaard_1996} and subsequent works by \citet{Yokoi_1999,Wang_2019,Teymourtash_2015,Bush_2006}. 

In the type I jump, the water continues to flow away from the disk after crossing the jump boundary. However, with greater submersion of the disk, flow pattern changes, resulting in water flowing back onto the top of the disk (type IIa jump). We observed that the type of jump coincides with the disk’s ability to float. The disk would only float when a fully developed type I jump, similar to the one depicted in Figure~\ref{fig:parametric_jumps}b, was present.


\section{Boundary jump}

\begin{figure}
\begin{center}
    \includegraphics[width=1\linewidth]{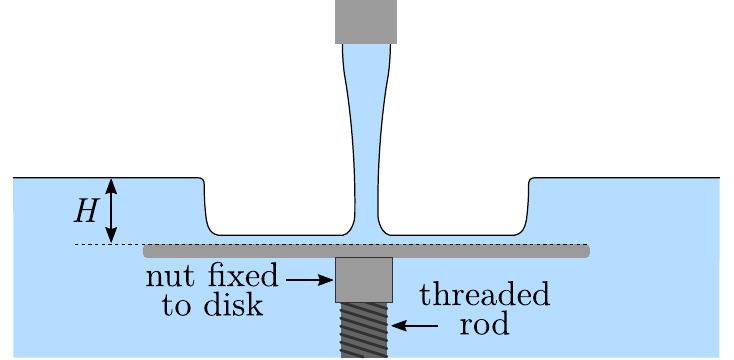}
\end{center}
\caption{
Experimental setup used to measure the critical depth $H_\textrm{crit}$. The disk's bottom was attached to a nut that could be moved along a vertical, threaded rod, affixed to the bottom of a large vessel. This arrangement enabled variation in the disk's submersion depth $H$.}
\label{fig:disk_on_stick_setup}
\end{figure}

As the first step, we focused on the flow dynamics, isolating them from the floating dynamics.
We achieved this by fixing the disk to the bottom of the liquid container using a variable-height pillar as in Figure~\ref{fig:disk_on_stick_setup}, which allowed us to precisely control the submersion depth of the disk.
Changing this depth allowed us to achieve the two boundary jump regimes, type I and type IIa. The nomenclature distinguishing between type IIa and IIb flows was proposed by \citet{Bush_2006}, where type IIb is associated with two layers of recirculation. However, in our observations of the boundary jump, such patterns did not appear.
As shown in the inset of Figure~\ref{fig:parametric_jumps}b, when the submersion depth was small, water on the surface flowed outward.
With an increase in the submersion depth, past a critical depth $H_\textrm{crit}$, a region of recirculation appeared near the edge of the disk (Figure~\ref{fig:parametric_jumps}c), causing the water to flow back onto the top surface of the disk.

\subsection{Scaling analysis}
The parameters of hydraulic jumps, including the radius or average velocity, can be modeled using scaling laws \citep{Bohr_1993, Duchesne_2014, Bhagat_2018, Baayoun_2022}. In line with these methodologies, we develop a model for $H_\textrm{crit}$ which marks the transition between type I and type IIa boundary jump.
Consider a jet impacting the centre of the disk (cf. Figure~\ref{fig:parametric_jumps}), where the impact generates a supercritical flow layer across the disk surface.
For the typical parameters used in our experiments ($Q = 100 \, \mathrm{ml/s}$ and cross-sectionally averaged jet velocity $U = 2 \, \mathrm{m/s}$), the influence of gravity on the flow pattern before the jump is negligible \citep{Wang_2019}, making the solution presented by \citet{Watson_1964} applicable.
This allows us to estimate the radius at which the boundary layer becomes fully developed. In most of our measurements, this radius remains smaller than $R$, consequently, near the disk edge, the flow adheres to the properties of a viscous boundary layer.

The fluid then interacts with the adjacent water, which, due to its significantly larger volume, possesses greater inertia. Consequently, the fluid's velocity near the edge reduces to a subcritical level, $u_{\textrm{low}}$. This deceleration results in the elevation of the water's surface to a height $H$, indicating a conversion of the fluid's kinetic energy into potential energy.

To establish the condition for the transition between the flow types, we examine the point on the edge of the disk where supercritical flow transitions to subcritical flow. We rely on the principle that the momentum flow must balance the pressure difference across the jump \citep{Rayleigh_1914, Watson_1964}. Denoting $u(z)$ and $h$ as a radial velocity and thickness of supercritical layer at the edge of the supercritical region the momentum balance takes form 
\begin{equation}
   \rho\, \overline{u(z)^2}\; h - \rho\, \overline{u_{\textrm{low}}^2}\; H = \frac{1}{2} \rho g \left( H^2 - h^2 \right),
    \label{eqn:momentum}
\end{equation}
where $g$ is the gravitational acceleration, $\rho$ is the water density, and the bar denotes a z-averaged quantity.
Guided by our experimental observations, we assume that 
$\overline{u(z)^2\vphantom{u_{\textrm{low}}^2}}
\gg 
\overline{u_{\textrm{low}}^2}$
and that $h\ll H$. 
We can quantify this intuition by considering the supercritical flow model developed by \citet{Watson_1964}, which, for experimentally relevant conditions, yields typical values of approximately $h \sim 0.1\,\textrm{mm}$ compared to $H \sim 5\,\textrm{mm}$. Using the conservation of mass, we can estimate the ratio 
$\overline{u_{\textrm{low}}\vphantom{u(z)}} 
/ \overline{u(z)}$, which suggests a typical value on the order of $10^{-2}$.
We proceed by rescaling the radial velocity using the surface velocity, $u_{\textrm{sup}}$, yielding $u(z) = u_{\textrm{sup}} \, \tilde{u}(z)$, where $\tilde{u}(z)$ is dimensionless. Similarly, we scale $z$ by the thickness of the layer, $z = h \zeta$.  
With these rescalings, the average squared velocity can be expressed as  
\begin{equation}  
    \overline{u(z)^2} = u_{\textrm{sup}}^2 \frac{1}{h} \int_0^1 \tilde{u}(\zeta)^2 h \, \textrm{d}\zeta = u_{\textrm{sup}}^2 \alpha_2,  
\end{equation}  
where $\alpha_2$ is an integration constant.  
The momentum balance then reduces to  
\begin{equation}  
    u_{\textrm{sup}}^2 h \alpha_2 = \frac{1}{2} g H^2,  
    \label{eqn:energy}  
\end{equation}  
which can be further rewritten using the mass conservation relation $Q = 2 \pi R h u_{\textrm{sup}} \alpha_1$, where $\alpha_1 = \int_0^1 \tilde{u}(\zeta)\, \textrm{d}\zeta$. Substituting this into the momentum balance gives  
\begin{equation}  
    \frac{Q^3}{8 \pi^3 R^3} \frac{1}{u_{\textrm{sup}} h^2} \frac{\alpha_2}{\alpha_1^3} = \frac{1}{2} g H^2.  
    \label{eqn:energy_ugly}  
\end{equation}
In order to derive the expression for $u_{\textrm{sup}}$, we employ an analogy with the classical hydraulic jump analysis of \citet{Watson_1964}. Under the assumptions of a fully developed viscous layer at the disk's edge and a sufficiently small jet radius $a$, we obtain the relation
\begin{equation}
    u_{\textrm{sup}} h^2 = \frac{\mu}{\rho} R \alpha_3.
    \label{eqn:balance}
\end{equation}
Finally, the unknown quantity $u_{\textrm{sup}} h^2$ can be eliminated from~\eqref{eqn:energy_ugly}. Since the dynamics of the liquid at the edge of the disk differ from those associated with the hydraulic jump studied in the literature, the values of $\alpha_{1,2,3}$ remain undetermined. However, we can still derive the desired scaling law for $H^2$,
\begin{equation}  
    H^2 \sim \frac{Q^3 \rho}{4 \pi^3 R^4 \mu g}.  
    \label{eqn:hsquared_scaling}  
\end{equation}  
We define the characteristic height $H^*$ as  
\begin{equation}  
    H^* = \sqrt{\frac{Q^3 \rho}{4 \pi^3 g \mu R^4}},  
    \label{eqn:dimlessH}  
\end{equation}  
which serves as a natural scale for $H_\textrm{crit}$.
By expressing $H_\textrm{crit}$ in units of $H^*$, we introduce the dimensionless prefactor $\xi$
\begin{equation}
    H_\textrm{crit} = \xi H^*,
    \label{eqn:linear}
\end{equation}
whose value was determined experimentally as described below.

\subsection{Experiments}\label{section:2Experiments}

\begin{figure*}
\begin{center}
    \includegraphics[width=1\linewidth]{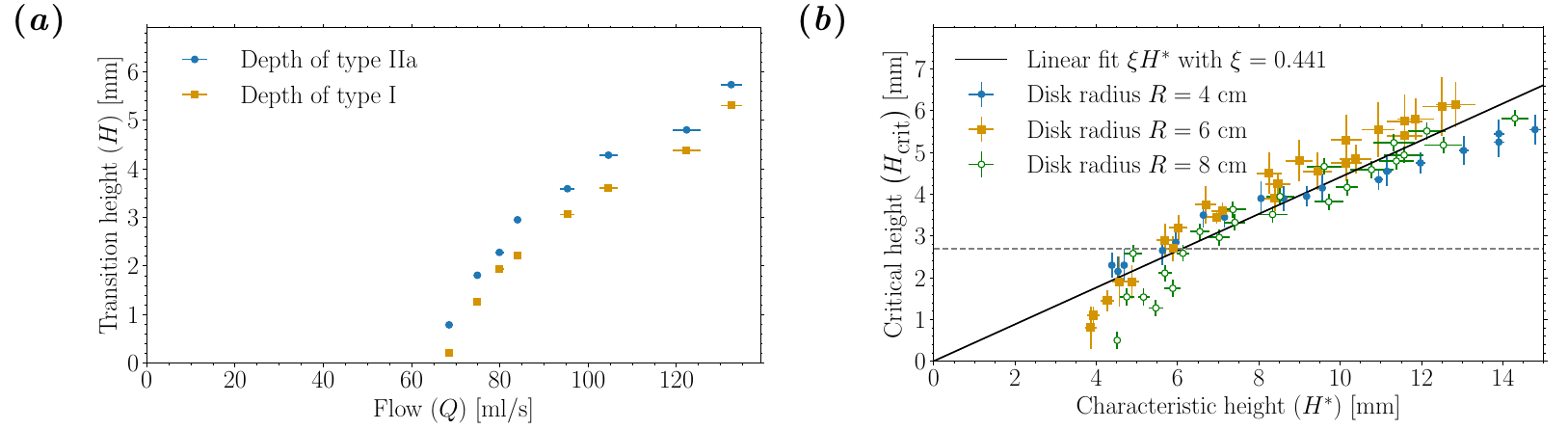}
\end{center}
\caption{(a) Measurement series for a disk with $R = 6 \, \mathrm{cm}$, varying $Q$ while keeping the nozzle height constant, resulting in a variable jet radius (and $U$). 
For each value of $Q$, the disk was gradually lowered, increasing $H$. Initially, only a type I jump was observed along the entire circumference. At a certain depth, the type IIa jump appeared along part of the disk edge (orange, square symbols). Eventually, the type II jump appeared around the entire circumferece (blue, round symbols). From these observations, $H_{\mathrm{crit}}$ was computed as the average value of the transition depths, with the measurement uncertainty taken as the difference between these values.
(b) A master curve showing $H_{\mathrm{crit}}$ against $H^*$, with different symbols showing disk radii. 
The dashed line indicates the capillary length ($2.7 \, \mathrm{mm}$). For depths above the capillary length, a linear fit (solid line) gives prefactor
$\xi = 0.441 \pm 0.007$.}
\label{fig:disk_on_stick_measurements}
\end{figure*}

To measure the transition depth accurately, the disk's position relative to the water surface $H$ was controlled using a nut on a vertical threaded rod fixed to the vessel bottom (Figure~\ref{fig:disk_on_stick_setup}). Turning the disk by $10^{\circ}$ adjusted its height by $54\,\mathrm{\mu m}$, allowing for $8\,\mathrm{mm}$ of vertical travel. First, the height was calibrated to the water surface by adjusting it until the meniscus was minimized on a dry disk. Then, the submersion depth was determined by measuring the rotation angle relative to the neutral meniscus position.

The jet was flowing out of a straight, $1\,\textrm{m}$ long, vertical pipe. This setup allowed for flow laminarisation and produced a smooth stream impacting the disk. $Q$ was regulated by a valve and quantified using a flow meter. The width of the stream was varied by changing the nozzle radius and the distance between the nozzle and the disk. The jet radius was experimentally determined by photographing the jet from the side. After enhancing the contrast of the photo, computer image analysis was employed to identify the edges and subsequently measure the jet radius $a$. For each measurement, 5 photos were taken, and the results were averaged. The standard deviation of the mean value of radius was around $1\%$. 

Knowing $Q$ and $a$ the cross-sectionally averaged velocity was found using the relation $Q = \pi a^2 U$. Our experiments revealed that the formation and shape of the boundary jump depend strongly on the downstream boundary condition, a sensitivity also noted by \citet{Bush_2006} in the case of hydraulic jumps.
We utilised a large cylindrical vessel, measuring $31\;\textrm{cm}$ in height and $50\;\textrm{cm}$ in diameter, which was carefully levelled.
This ensured minimal water height deviations at the container's edge (less than $3\;\textrm{mm}$). 
It was difficult to achieve better accuracy due to surface tension effects at the edge of the container.
As water was overflowing the edge of the vessel, 
$H$ remained constant during the experiment. 

The setup was used to determine the submersion depth of the disk at which the type I and type IIa jumps occurred across various values of $Q$, $U$ and $R$. We used disks with radii $R=4,6,8\;\textrm{cm}$. Each series was conducted while maintaining a constant nozzle height above the water (note that $U$ varied in these series). To vary $U$ and $Q$ independently, we used different nozzle diameters and heights above the water in different series, achieving a range of $U$ between $1$ m/s and $3$ m/s.

Unfortunately the transition between different jump types did not occur simultaneously across the entire disk. 
Instead, during these transitions, we observed a type I jump on some parts of the disk and a type IIa jump on others. 
For each measurement, we recorded the greatest depth for which type I jump was fully developed and the shallowest depth for which type IIa jump was fully developed. 
The experimentally measured critical height, denoted as $H_\mathrm{crit}$, was calculated as the average of these extreme values.
An illustrative measurement series for jump height is depicted in Figure~\ref{fig:disk_on_stick_measurements}a. Utilising the presented scaling analysis \eqref{eqn:linear}, we were able to combine all measurement series into a single master curve, as shown in Figure~\ref{fig:disk_on_stick_measurements}b.
    
Within the studied parameter range, $H_\textrm{crit}$ adheres to the predicted linear relationship \eqref{eqn:linear}.
A deviation from linear dependency can be noticed for $H_\textrm{crit} \lesssim 2.5 \,\mathrm{mm}$ which aligns with the capillary length of water $l_{\textrm{cap}} \approx 2.7 \,\mathrm{mm}$. 
This implies that surface tension phenomena might influence the value of $H_\textrm{crit}$ at smaller $H^*$ values.

\section{Unsinkable disk}

The phenomenon, referred to as the \emph{unsinkable disk}, emerges when a vertically directed jet impinges upon the centre of a freely floating metallic disk. 
Under specific conditions, a disk that would ordinarily submerge can instead achieve a floating equilibrium.
Remarkably, this phenomenon exhibits self-stabilising characteristics; when the disk is disturbed, either by a change in inclination or position, it returns to a stable configuration.
Video demonstration of the effect can be seen at \url{https://youtu.be/as0wRQj1Zws}.

Initial observations revealed that the phenomenon of the unsinkable disk manifests exclusively under type I jump conditions.
Conversely, the disk sinks when a type IIa jump is observed.
This insight suggests that the system parameters needed for the disk to float are congruent with those that give rise to a type I jump.
This correlation can be intuitively understood. 
In the case of a type IIa jump, a water vortex forms (as depicted in Figure~\ref{fig:parametric_jumps}c) which redirects the water back onto the disk. 
This change diminishes the effective buoyant force, ultimately causing the disk to sink.

\subsection{Force balance}

For the disk to achieve a state of equilibrium, in which it floats, the net force exerted upon it must be zero,
\begin{equation}
    0 = F_{\textrm{b}}-F_{\textrm{j}}-F_{\textrm{g}} 
    \label{eqn:forces}.
\end{equation}
The forces contributing to this equilibrium include: the force generated by the impinging jet, $F_{\textrm{j}} = \rho Q^2/\left(\pi a^2\right)$\citep{Stephens_Applied_Mechanics}; the gravitational force acting on the disk (with mass $m$), $F_{\textrm{g}} = g m$; and buoyancy force $F_{\textrm{b}}$ arising from both the volume of water displaced by the disk itself $V$ and the additional volume of air created by water displaced by the jet.
Due to negligible thickness of the water layer the latter volume, is a function of the disk's floating depth $H_\textrm{float}$ and its surface area $S = \pi R^2$, thus
\begin{equation}
    F_{\textrm{b}} = \rho g (V+H_\textrm{float}S).
    \label{eqn:buoyancy_force}
\end{equation}
With the effective mass of the disk as $m_{\textrm{eff}}=m-\rho V$, the floating depth is
\begin{equation}
    H_\textrm{float} = \frac{m_{\textrm{eff}}}{\rho S} + \frac{Q^2}{g S \pi a^2}.
    \label{eqn:floating_height}
\end{equation}

The disk can float only if the jet has sufficient momentum to displace the necessary amount of water -- i.e., if a type I jump occurs. Conversely, when a type IIa jump forms, water flows onto the disk, increasing the pressure and thereby reducing $ F_{\textrm{b}} $ compared to the value predicted by Equation~\eqref{eqn:buoyancy_force}. As a result, flotation requires a greater $ H_\textrm{float} $, which, in turn, increases the amount of water on top of the disk, further decreasing $ F_{\textrm{b}} $. This feedback mechanism ultimately leads to sinking. Thus $H_\mathrm{crit} > H_\mathrm{float}$ is the criterion for the disk to remain afloat, yielding 
\begin{equation}
\xi \sqrt{\frac{Q^3 \rho }{4 \pi \mu g}} > \frac{m_{\textrm{eff}}}{\rho} + \frac{Q^2}{g \pi a^2}.
\label{eqn:floating_criterion_raw}
\end{equation}
Using the jet's Reynolds ($\reynolds$) and Froude ($\froude$) numbers, defined as
\begin{equation}
\reynolds = \frac{2Q\rho}{\pi a \mu} \qquad \text{and} \qquad \froude = \frac{Q}{\pi a^2\sqrt{ag}},
\label{eqn:re_fr}
\end{equation}
Equation~\eqref{eqn:floating_criterion_raw} simplifies to
\begin{equation}
\frac{\xi}{\sqrt{8}} \sqrt{\reynolds} > \frac{1}{\froude} \frac{m_\textrm{eff}/\rho}{\pi a^3} + \froude.
\label{eqn:floating_criterion_three}
\end{equation}
Three dimensionless quantities appear in Equation~\eqref{eqn:floating_criterion_three}: $\reynolds$, $\froude$, and $\omega^2 = m_\textrm{eff} / (\pi a^3 \rho)$. Through algebraic manipulation, the number of dimensionless constants can be reduced to two: $\sqrt{\reynolds}/\omega$ and $\froude/\omega$, yielding the floating criterion
\begin{equation}
\frac{\xi}{\sqrt{8}} \frac{\sqrt{\reynolds}}{\omega} > \frac{\omega}{\froude}+\frac{\froude}{\omega}.
\label{eqn:floating_criterion}
\end{equation}

In our theoretical calculations we used density of water $\rho = 997 \, \textrm{kg}/\textrm{m}^3$, and its kinematic viscosity $\mu/\rho=0.9\times10^{-6}\,\textrm{m}^2/\textrm{s}$ as reported by \citet{Kestin_1978}.

\subsection{Experiments}
\begin{figure*}[t]
    \begin{center}
    \includegraphics[width=1\linewidth]{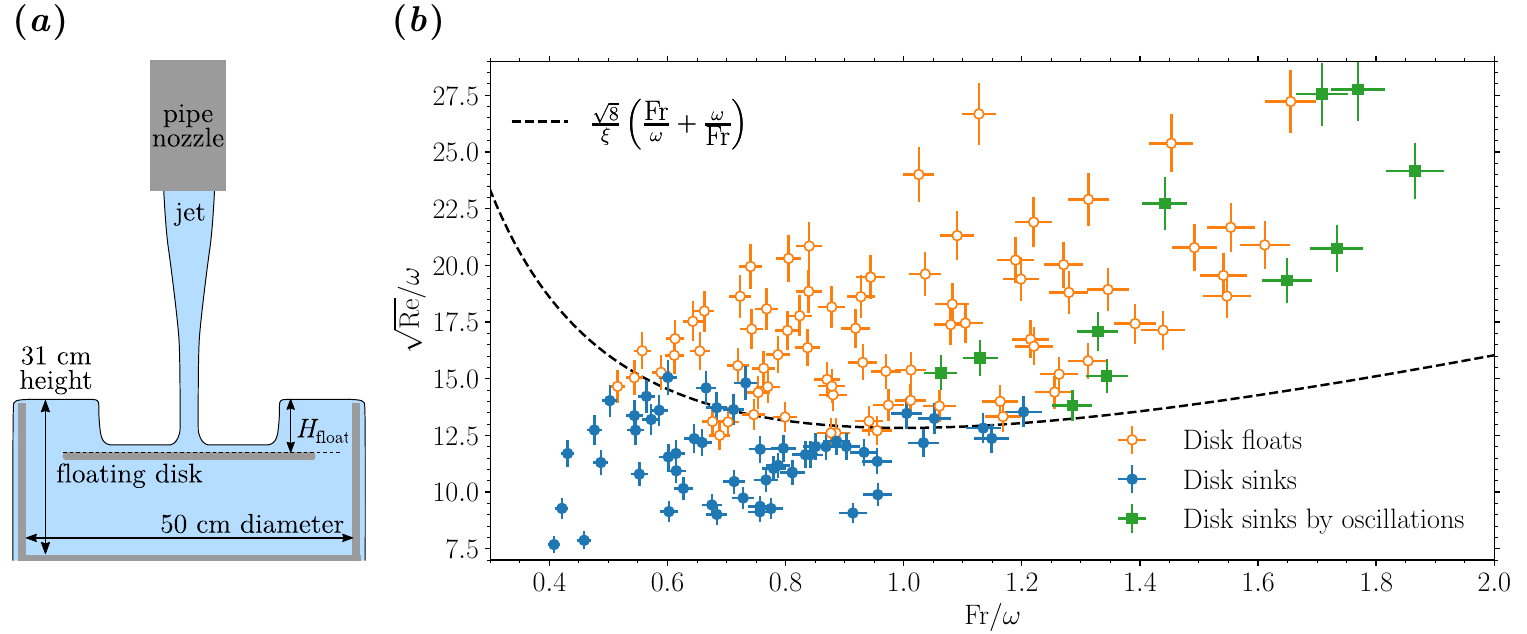}   
    \end{center}
    \caption{  
    (a) Experimental setup used to investigate whether the disk would float or sink. The flow rate $Q$ and jet radius $a$ were independently controlled by adjusting the nozzle position and size. These parameters were used to compute the jet velocity $U$.  
    (b) 
    Results from the floating/sinking experiments presented in dimensionless form, where $\froude$ and $\reynolds$ are the jet's Froude and Reynolds numbers defined in Equation~\eqref{eqn:re_fr}, and $\omega$ is the dimensionless effective mass of the disk, defined by $\omega^2=m_\textrm{eff}/(\pi a^3 \rho)$. Experimental data (symbols) show good quantitative agreement with the theoretical prediction of the float/sink boundary (dashed black line), except for measurements labeled "Disk sinks by oscillations" (green symbols), in which sinking was triggered by increasing tilt oscillations.} 
    \label{fig:float_nofloat_experiments}
\end{figure*}

To test the necessary conditions for the disk to float, a similar experimental setup to the one described in the Section~\ref{section:2Experiments} was used. 
This time, the disk was unrestricted and free to move, as shown in Figure~\ref{fig:float_nofloat_experiments}a. 
The disks used in the experiment were carefully positioned beneath the jet and subsequently released. 
We observed that the disks sank almost immediately when $Q$ and $U$ were insufficient. 
The disk was considered to be floating if it remained on the surface for at least 15 seconds.

The disks used in the experiments were $0.1\;\mathrm{mm}$ thick and made of aluminium, with radii of $R=4,5,7\;\textrm{cm}$, masses of $m=11.7\pm0.1,30.7\pm0.1,50.9\pm0.2\;\textrm{g}$ and volumes of $V = 2.77\pm0.05,14.92\pm0.05,12.16\pm0.05\;\textrm{ml}$ respectively.
Some disks had their weight and volume changed by copper washers glued to their underside. The additional buoyancy and weight resulting from these modifications have been taken into account in the comparison with theoretical predictions. 

We studied $U$ in the range from $0.6\;\mathrm{m/s}$ to $2.5\;\mathrm{m/s}$ and focused on $Q$ close to the floating-sinking threshold. The results for each disk are available in the Supplemental Materials.
Comparison between the theoretical model and measurements is shown in
Figure~\ref{fig:float_nofloat_experiments}b. 
Generally when values of flow $Q$ were sufficiently high the disk would remain afloat.
When $Q$ was too low, the disk would sink deeper until the type I jump transitioned to type IIa, allowing water to flow onto the disk (cf. Figure \ref{fig:parametric_jumps}c). This, in turn, increased the floating depth and added more mass to the water atop the disk, further submerging it. As the process continued, the supercritical layer quickly disappeared, ultimately causing the disk to sink.
Observations proved that, in every case, transition to type IIa resulted in abrupt sinking. This aligns with the assumption used in theoretical predictions that the transition between type I and type IIa jumps determines the floating/sinking criterion.

The theoretical model aligns with most cases studied, but deviations appear at high $U$, marked as ``Disk sinks by oscillations''. While the theory predicts floating, in experiments the disk sank, by oscillating with increasing amplitude, lateral movement, and tilting. The jet force pushed the disk's lower side downward, causing water overflow and submersion. 
Although a detailed rotational stability analysis is beyond this paper's scope, in cases marked as ``Disk floats'', the disk consistently returned to a stable position after external disturbance. 
All data from experimental results presented above as well as codes responsible for making plots are available in a github repository \cite{our_repository}.

\section{Conclusions}
Our study demonstrates that directing a vertical jet onto a disk from above can prevent it from sinking. This effect is attributed to the boundary jump phenomenon, which displaces water from the top of the disk, generating an additional upward buoyancy force. We identified two distinct types of boundary jumps, type I and type IIa, and established that maintaining the disk's flotation is feasible under type I jump conditions.

The boundary jump differs from the classical hydraulic jump, as the position of the water jump is fixed by the geometry of the setup. In this case, we utilised a scaling law to determine the critical submersion depth at which the transition from type I to type IIa occurs. The prefactor
, $\xi = 0.441 \pm 0.007$, was determined through an independent experiment measuring the submersion depth corresponding to this critical transition.

The phenomenon of the floating disk was studied experimentally across a broad range of parameters, including the disk’s radius, jet flow, and jet radius. Results were compared against the theoretical model without parameter fitting, showing substantial agreement in nearly all cases examined.

Notably, at high jet flow and velocities, the disk exhibited signs of instability, characterised by progressively intensifying oscillations. This observation suggests that while the proposed model establishes the necessary conditions for disk flotation, it does not yet encompass all the sufficient conditions required for a comprehensive understanding of the system’s stability.

\begin{acknowledgments}
We extend our sincere gratitude to Rajesh Bhagat for his invaluable guidance on the theoretical aspects of this publication.

The research presented herein was conducted in part during the preparations of the Fenix Science Club's (Klub Naukowy Fenix) Polish team for the International Young Physicist Tournament (IYPT) 2022. The team comprised Jan Turczynowicz, Maciej Dąbkowski, Mikołaj Czarnecki, Igor Kumela, Rafał Bryl, with Łukasz Gładczuk and Radost Waszkiewicz serving as team leaders. 
This particular work addresses the problem \emph{4. Unsinkable Disk} posed at the IYPT 2022.
\end{acknowledgments}

\bibliography{sources}

\providecommand{\noopsort}[1]{}\providecommand{\singleletter}[1]{#1}%
\begin{thebibliography}{19}%
\makeatletter
\providecommand \@ifxundefined [1]{%
 \@ifx{#1\undefined}
}%
\providecommand \@ifnum [1]{%
 \ifnum #1\expandafter \@firstoftwo
 \else \expandafter \@secondoftwo
 \fi
}%
\providecommand \@ifx [1]{%
 \ifx #1\expandafter \@firstoftwo
 \else \expandafter \@secondoftwo
 \fi
}%
\providecommand \natexlab [1]{#1}%
\providecommand \enquote  [1]{``#1''}%
\providecommand \bibnamefont  [1]{#1}%
\providecommand \bibfnamefont [1]{#1}%
\providecommand \citenamefont [1]{#1}%
\providecommand \href@noop [0]{\@secondoftwo}%
\providecommand \href [0]{\begingroup \@sanitize@url \@href}%
\providecommand \@href[1]{\@@startlink{#1}\@@href}%
\providecommand \@@href[1]{\endgroup#1\@@endlink}%
\providecommand \@sanitize@url [0]{\catcode `\\12\catcode `\$12\catcode
  `\&12\catcode `\#12\catcode `\^12\catcode `\_12\catcode `\%12\relax}%
\providecommand \@@startlink[1]{}%
\providecommand \@@endlink[0]{}%
\providecommand \url  [0]{\begingroup\@sanitize@url \@url }%
\providecommand \@url [1]{\endgroup\@href {#1}{\urlprefix }}%
\providecommand \urlprefix  [0]{URL }%
\providecommand \Eprint [0]{\href }%
\providecommand \doibase [0]{https://doi.org/}%
\providecommand \selectlanguage [0]{\@gobble}%
\providecommand \bibinfo  [0]{\@secondoftwo}%
\providecommand \bibfield  [0]{\@secondoftwo}%
\providecommand \translation [1]{[#1]}%
\providecommand \BibitemOpen [0]{}%
\providecommand \bibitemStop [0]{}%
\providecommand \bibitemNoStop [0]{.\EOS\space}%
\providecommand \EOS [0]{\spacefactor3000\relax}%
\providecommand \BibitemShut  [1]{\csname bibitem#1\endcsname}%
\let\auto@bib@innerbib\@empty
\bibitem [{\citenamefont {{Lord Rayleigh}}(1914)}]{Rayleigh_1914}%
  \BibitemOpen
  \bibfield  {author} {\bibinfo {author} {\bibnamefont {{Lord Rayleigh}}},\
  }\bibfield  {title} {\bibinfo {title} {On the theory of long waves and
  bores},\ }\href {https://doi.org/10.1098/rspa.1914.0055} {\bibfield
  {journal} {\bibinfo  {journal} {Proc. R. Soc. Lond. A}\ }\textbf {\bibinfo
  {volume} {90}},\ \bibinfo {pages} {324} (\bibinfo {year} {1914})}\BibitemShut
  {NoStop}%
\bibitem [{\citenamefont {Ivanova}\ and\ \citenamefont
  {Gavrilyuk}(2019)}]{Ivanova_2019}%
  \BibitemOpen
  \bibfield  {author} {\bibinfo {author} {\bibfnamefont {K.~A.}\ \bibnamefont
  {Ivanova}}\ and\ \bibinfo {author} {\bibfnamefont {S.~L.}\ \bibnamefont
  {Gavrilyuk}},\ }\bibfield  {title} {\bibinfo {title} {Structure of the
  hydraulic jump in convergent radial flows},\ }\href
  {https://doi.org/10.1017/jfm.2018.901} {\bibfield  {journal} {\bibinfo
  {journal} {J. Fluid. Mech.}\ }\textbf {\bibinfo {volume} {860}},\ \bibinfo
  {pages} {441–464} (\bibinfo {year} {2019})}\BibitemShut {NoStop}%
\bibitem [{\citenamefont {Teymourtash}\ and\ \citenamefont
  {Mokhlesi}(2015)}]{Teymourtash_2015}%
  \BibitemOpen
  \bibfield  {author} {\bibinfo {author} {\bibfnamefont {A.~R.}\ \bibnamefont
  {Teymourtash}}\ and\ \bibinfo {author} {\bibfnamefont {M.}~\bibnamefont
  {Mokhlesi}},\ }\bibfield  {title} {\bibinfo {title} {Experimental
  investigation of stationary and rotational structures in non-circular
  hydraulic jumps},\ }\href {https://doi.org/10.1017/jfm.2014.646} {\bibfield
  {journal} {\bibinfo  {journal} {J. Fluid. Mech.}\ }\textbf {\bibinfo {volume}
  {762}},\ \bibinfo {pages} {344–360} (\bibinfo {year} {2015})}\BibitemShut
  {NoStop}%
\bibitem [{\citenamefont {Wang}\ and\ \citenamefont
  {Khayat}(2019)}]{Wang_2019}%
  \BibitemOpen
  \bibfield  {author} {\bibinfo {author} {\bibfnamefont {Y.}~\bibnamefont
  {Wang}}\ and\ \bibinfo {author} {\bibfnamefont {R.~E.}\ \bibnamefont
  {Khayat}},\ }\bibfield  {title} {\bibinfo {title} {The role of gravity in the
  prediction of the circular hydraulic jump radius for high-viscosity
  liquids},\ }\href {https://doi.org/10.1017/jfm.2018.941} {\bibfield
  {journal} {\bibinfo  {journal} {J. Fluid. Mech.}\ }\textbf {\bibinfo {volume}
  {862}},\ \bibinfo {pages} {128–161} (\bibinfo {year} {2019})}\BibitemShut
  {NoStop}%
\bibitem [{\citenamefont {Wang}\ and\ \citenamefont
  {Khayat}(2018)}]{Wang_2018}%
  \BibitemOpen
  \bibfield  {author} {\bibinfo {author} {\bibfnamefont {Y.}~\bibnamefont
  {Wang}}\ and\ \bibinfo {author} {\bibfnamefont {R.~E.}\ \bibnamefont
  {Khayat}},\ }\bibfield  {title} {\bibinfo {title} {Impinging jet flow and
  hydraulic jump on a rotating disk},\ }\href
  {https://doi.org/10.1017/jfm.2018.43} {\bibfield  {journal} {\bibinfo
  {journal} {J. Fluid. Mech.}\ }\textbf {\bibinfo {volume} {839}},\ \bibinfo
  {pages} {525–560} (\bibinfo {year} {2018})}\BibitemShut {NoStop}%
\bibitem [{\citenamefont {Baayoun}\ \emph {et~al.}(2022)\citenamefont
  {Baayoun}, \citenamefont {Khayat},\ and\ \citenamefont
  {Wang}}]{Baayoun_2022}%
  \BibitemOpen
  \bibfield  {author} {\bibinfo {author} {\bibfnamefont {A.}~\bibnamefont
  {Baayoun}}, \bibinfo {author} {\bibfnamefont {R.~E.}\ \bibnamefont
  {Khayat}},\ and\ \bibinfo {author} {\bibfnamefont {Y.}~\bibnamefont {Wang}},\
  }\bibfield  {title} {\bibinfo {title} {The transient spread of a circular
  liquid jet and hydraulic jump formation},\ }\href
  {https://doi.org/10.1017/jfm.2022.670} {\bibfield  {journal} {\bibinfo
  {journal} {J. Fluid. Mech.}\ }\textbf {\bibinfo {volume} {947}},\ \bibinfo
  {pages} {A34} (\bibinfo {year} {2022})}\BibitemShut {NoStop}%
\bibitem [{\citenamefont {Duchesne}\ and\ \citenamefont
  {Limat}(2022)}]{Duchesne_2022}%
  \BibitemOpen
  \bibfield  {author} {\bibinfo {author} {\bibfnamefont {A.}~\bibnamefont
  {Duchesne}}\ and\ \bibinfo {author} {\bibfnamefont {L.}~\bibnamefont
  {Limat}},\ }\bibfield  {title} {\bibinfo {title} {Circular hydraulic jumps:
  where does surface tension matter?},\ }\href
  {https://doi.org/10.1017/jfm.2022.136} {\bibfield  {journal} {\bibinfo
  {journal} {J. Fluid. Mech.}\ }\textbf {\bibinfo {volume} {937}},\ \bibinfo
  {pages} {R2} (\bibinfo {year} {2022})}\BibitemShut {NoStop}%
\bibitem [{\citenamefont {Watson}(1964)}]{Watson_1964}%
  \BibitemOpen
  \bibfield  {author} {\bibinfo {author} {\bibfnamefont {E.~J.}\ \bibnamefont
  {Watson}},\ }\bibfield  {title} {\bibinfo {title} {The radial spread of a
  liquid jet over a horizontal plane},\ }\href
  {https://doi.org/10.1017/S0022112064001367} {\bibfield  {journal} {\bibinfo
  {journal} {J. Fluid. Mech.}\ }\textbf {\bibinfo {volume} {20}},\ \bibinfo
  {pages} {481–499} (\bibinfo {year} {1964})}\BibitemShut {NoStop}%
\bibitem [{\citenamefont {Bohr}\ \emph {et~al.}(1993)\citenamefont {Bohr},
  \citenamefont {Dimon},\ and\ \citenamefont {Putkaradze}}]{Bohr_1993}%
  \BibitemOpen
  \bibfield  {author} {\bibinfo {author} {\bibfnamefont {T.}~\bibnamefont
  {Bohr}}, \bibinfo {author} {\bibfnamefont {P.}~\bibnamefont {Dimon}},\ and\
  \bibinfo {author} {\bibfnamefont {V.}~\bibnamefont {Putkaradze}},\ }\bibfield
   {title} {\bibinfo {title} {Shallow-water approach to the circular hydraulic
  jump},\ }\href {https://doi.org/10.1017/S0022112093002289} {\bibfield
  {journal} {\bibinfo  {journal} {J. Fluid. Mech.}\ }\textbf {\bibinfo {volume}
  {254}},\ \bibinfo {pages} {635–648} (\bibinfo {year} {1993})}\BibitemShut
  {NoStop}%
\bibitem [{\citenamefont {Bohr}\ and\ \citenamefont
  {Scheichl}(2021)}]{Bohr_2021}%
  \BibitemOpen
  \bibfield  {author} {\bibinfo {author} {\bibfnamefont {T.}~\bibnamefont
  {Bohr}}\ and\ \bibinfo {author} {\bibfnamefont {B.}~\bibnamefont
  {Scheichl}},\ }\bibfield  {title} {\bibinfo {title} {Surface tension and
  energy conservation in a moving fluid},\ }\href
  {https://doi.org/10.1103/PhysRevFluids.6.L052001} {\bibfield  {journal}
  {\bibinfo  {journal} {Phys. Rev. Fluids}\ }\textbf {\bibinfo {volume} {6}},\
  \bibinfo {pages} {L052001} (\bibinfo {year} {2021})}\BibitemShut {NoStop}%
\bibitem [{\citenamefont {Duchesne}\ \emph {et~al.}(2019)\citenamefont
  {Duchesne}, \citenamefont {Andersen},\ and\ \citenamefont
  {Bohr}}]{Duchesne_2019}%
  \BibitemOpen
  \bibfield  {author} {\bibinfo {author} {\bibfnamefont {A.}~\bibnamefont
  {Duchesne}}, \bibinfo {author} {\bibfnamefont {A.}~\bibnamefont {Andersen}},\
  and\ \bibinfo {author} {\bibfnamefont {T.}~\bibnamefont {Bohr}},\ }\bibfield
  {title} {\bibinfo {title} {Surface tension and the origin of the circular
  hydraulic jump in a thin liquid film},\ }\href
  {https://doi.org/10.1103/PhysRevFluids.4.084001} {\bibfield  {journal}
  {\bibinfo  {journal} {Phys. Rev. Fluids}\ }\textbf {\bibinfo {volume} {4}},\
  \bibinfo {pages} {084001} (\bibinfo {year} {2019})}\BibitemShut {NoStop}%
\bibitem [{\citenamefont {Ellegaard}\ \emph {et~al.}(1996)\citenamefont
  {Ellegaard}, \citenamefont {Hansen}, \citenamefont {Haaning},\ and\
  \citenamefont {Bohr}}]{Ellegaard_1996}%
  \BibitemOpen
  \bibfield  {author} {\bibinfo {author} {\bibfnamefont {C.}~\bibnamefont
  {Ellegaard}}, \bibinfo {author} {\bibfnamefont {A.~E.}\ \bibnamefont
  {Hansen}}, \bibinfo {author} {\bibfnamefont {A.}~\bibnamefont {Haaning}},\
  and\ \bibinfo {author} {\bibfnamefont {T.}~\bibnamefont {Bohr}},\ }\bibfield
  {title} {\bibinfo {title} {Experimental results on flow separation and
  transitions in the circular hydraulic jump},\ }\href
  {https://doi.org/10.1088/0031-8949/1996/T67/021} {\bibfield  {journal}
  {\bibinfo  {journal} {Phys. Scr.}\ }\textbf {\bibinfo {volume} {1996}},\
  \bibinfo {pages} {105} (\bibinfo {year} {1996})}\BibitemShut {NoStop}%
\bibitem [{\citenamefont {Yokoi}\ and\ \citenamefont
  {Xiao}(1999)}]{Yokoi_1999}%
  \BibitemOpen
  \bibfield  {author} {\bibinfo {author} {\bibfnamefont {K.}~\bibnamefont
  {Yokoi}}\ and\ \bibinfo {author} {\bibfnamefont {F.}~\bibnamefont {Xiao}},\
  }\bibfield  {title} {\bibinfo {title} {A numerical study of the transition in
  the circular hydraulic jump},\ }\href
  {https://doi.org/https://doi.org/10.1016/S0375-9601(99)00287-X} {\bibfield
  {journal} {\bibinfo  {journal} {Phys. Lett. A}\ }\textbf {\bibinfo {volume}
  {257}},\ \bibinfo {pages} {153} (\bibinfo {year} {1999})}\BibitemShut
  {NoStop}%
\bibitem [{\citenamefont {Bush}\ \emph {et~al.}(2006)\citenamefont {Bush},
  \citenamefont {Aristoff},\ and\ \citenamefont {Hosoi}}]{Bush_2006}%
  \BibitemOpen
  \bibfield  {author} {\bibinfo {author} {\bibfnamefont {J.~W.~M.}\
  \bibnamefont {Bush}}, \bibinfo {author} {\bibfnamefont {J.~M.}\ \bibnamefont
  {Aristoff}},\ and\ \bibinfo {author} {\bibfnamefont {A.~E.}\ \bibnamefont
  {Hosoi}},\ }\bibfield  {title} {\bibinfo {title} {An experimental
  investigation of the stability of the circular hydraulic jump},\ }\href
  {https://doi.org/10.1017/s0022112006009839} {\bibfield  {journal} {\bibinfo
  {journal} {J. Fluid. Mech.}\ }\textbf {\bibinfo {volume} {558}},\ \bibinfo
  {pages} {33–52} (\bibinfo {year} {2006})}\BibitemShut {NoStop}%
\bibitem [{\citenamefont {Duchesne}\ \emph {et~al.}(2014)\citenamefont
  {Duchesne}, \citenamefont {Lebon},\ and\ \citenamefont
  {Limat}}]{Duchesne_2014}%
  \BibitemOpen
  \bibfield  {author} {\bibinfo {author} {\bibfnamefont {A.}~\bibnamefont
  {Duchesne}}, \bibinfo {author} {\bibfnamefont {L.}~\bibnamefont {Lebon}},\
  and\ \bibinfo {author} {\bibfnamefont {L.}~\bibnamefont {Limat}},\ }\bibfield
   {title} {\bibinfo {title} {Constant froude number in a circular hydraulic
  jump and its implication on the jump radius selection},\ }\href
  {https://doi.org/10.1209/0295-5075/107/54002} {\bibfield  {journal} {\bibinfo
   {journal} {Europhys. Lett.}\ }\textbf {\bibinfo {volume} {107}},\ \bibinfo
  {pages} {54002} (\bibinfo {year} {2014})}\BibitemShut {NoStop}%
\bibitem [{\citenamefont {Bhagat}\ \emph {et~al.}(2018)\citenamefont {Bhagat},
  \citenamefont {Jha}, \citenamefont {Linden},\ and\ \citenamefont
  {Wilson}}]{Bhagat_2018}%
  \BibitemOpen
  \bibfield  {author} {\bibinfo {author} {\bibfnamefont {R.~K.}\ \bibnamefont
  {Bhagat}}, \bibinfo {author} {\bibfnamefont {N.~K.}\ \bibnamefont {Jha}},
  \bibinfo {author} {\bibfnamefont {P.~F.}\ \bibnamefont {Linden}},\ and\
  \bibinfo {author} {\bibfnamefont {D.~I.}\ \bibnamefont {Wilson}},\ }\bibfield
   {title} {\bibinfo {title} {On the origin of the circular hydraulic jump in a
  thin liquid film},\ }\href {https://doi.org/10.1017/jfm.2018.558} {\bibfield
  {journal} {\bibinfo  {journal} {J. Fluid. Mech.}\ }\textbf {\bibinfo {volume}
  {851}},\ \bibinfo {pages} {R5} (\bibinfo {year} {2018})}\BibitemShut
  {NoStop}%
\bibitem [{\citenamefont {Stephens}\ and\ \citenamefont
  {Ward}(1972)}]{Stephens_Applied_Mechanics}%
  \BibitemOpen
  \bibfield  {author} {\bibinfo {author} {\bibfnamefont {R.~C.}\ \bibnamefont
  {Stephens}}\ and\ \bibinfo {author} {\bibfnamefont {J.~J.}\ \bibnamefont
  {Ward}},\ }\href {https://doi.org/10.1007/978-1-349-00870-4_16} {\emph
  {\bibinfo {title} {Applied Mechanics}}}\ (\bibinfo  {publisher} {Macmillan
  Education UK},\ \bibinfo {address} {London},\ \bibinfo {year} {1972})\ p.\
  \bibinfo {pages} {170}\BibitemShut {NoStop}%
\bibitem [{\citenamefont {Kestin}\ \emph {et~al.}(1978)\citenamefont {Kestin},
  \citenamefont {Sokolov},\ and\ \citenamefont {Wakeham}}]{Kestin_1978}%
  \BibitemOpen
  \bibfield  {author} {\bibinfo {author} {\bibfnamefont {J.}~\bibnamefont
  {Kestin}}, \bibinfo {author} {\bibfnamefont {M.}~\bibnamefont {Sokolov}},\
  and\ \bibinfo {author} {\bibfnamefont {W.~A.}\ \bibnamefont {Wakeham}},\
  }\bibfield  {title} {\bibinfo {title} {Viscosity of liquid water in the range
  -8$^{\circ}$c to 150$^{\circ}$c},\ }\href {https://doi.org/10.1063/1.555581}
  {\bibfield  {journal} {\bibinfo  {journal} {J. Phys. Chem. Ref. Data}\
  }\textbf {\bibinfo {volume} {7}},\ \bibinfo {pages} {941} (\bibinfo {year}
  {1978})}\BibitemShut {NoStop}%
\bibitem [{\citenamefont {Turczynowicz}\ \emph {et~al.}()\citenamefont
  {Turczynowicz}, \citenamefont {Waszkiewicz},\ and\ \citenamefont
  {Gladczuk}}]{our_repository}%
  \BibitemOpen
  \bibfield  {author} {\bibinfo {author} {\bibfnamefont {J.}~\bibnamefont
  {Turczynowicz}}, \bibinfo {author} {\bibfnamefont {R.}~\bibnamefont
  {Waszkiewicz}},\ and\ \bibinfo {author} {\bibfnamefont {L.}~\bibnamefont
  {Gladczuk}},\ }\href {https://github.com/Fenix-Science-Club/unsinkable-disk}
  {\bibinfo {title} {github repository with data and python codes making
  plots}}\BibitemShut {NoStop}%
\end{thebibliography}%

\end{document}